\documentstyle[preprint,eqsecnum,aps,amsmath,epsfig]{revtex}
\newcommand{\psl}{\ooalign{\hfil/\hfil\crcr$P$}}
\newcommand{\ksl}{\ooalign{\hfil/\hfil\crcr$k$}}
\newcommand{\ssl}{\ooalign{\hfil/\hfil\crcr$S$}}
\renewcommand{\theequation}{\arabic{section}.\arabic{equation}}
\renewcommand{\thefigure}{\arabic{figure}}
\renewcommand{\thetable}{\arabic{section}.\arabic{table}}

\begin{document}

\draft
\title{\ \ \ \\
       \vspace{-2.0cm}
       \hfill SAGA-HE-139-98 \\
                           \ \\
%%\title{Structure functions in the polarized Drell-Yan processes \\
         Structure functions in the polarized Drell-Yan processes \\
         with spin-1/2 and spin-1 hadrons:
         \uppercase\expandafter{\romannumeral 2}. parton model}
%%         II. parton model}
\vspace{-0.3cm}
\author{S. Hino and S. Kumano \cite{byline}}
\address{Department of Physics, Saga University \\
         Saga, 840-8502, Japan}
%%\date{\today}
\date{Feb. 6, 1999}
\maketitle
\vspace{-0.5cm}
\begin{abstract}
We analyze the polarized Drell-Yan processes with spin-1/2 and spin-1
hadrons in a parton model. Quark and antiquark correlation functions are
expressed in terms of possible combinations of Lorentz vectors and
pseudovectors with the constrains of Hermiticity, parity conservation,
and time-reversal invariance. Then, we find tensor polarized
distributions for a spin-1 hadron. The naive parton model predicts
that there exist 19 structure functions. However, there are only
four or five non-vanishing structure functions, depending on whether
the cross section is integrated over the virtual-photon transverse
momentum $\vec Q_T$ or the limit $Q_T\rightarrow 0$ is taken.
One of the finite structure functions is related to the tensor polarized
distribution $b_1$, and it does not exist in the proton-proton
reactions. The vanishing structure functions should be associated with
higher-twist physics. The tensor distributions
can be measured by the quadrupole polarization measurements.
The Drell-Yan process has an advantage over the lepton reaction
in the sense that the antiquark tensor polarization could be
extracted rather easily.
\end{abstract}

\narrowtext

%%%%%%%%%%%%%%%%%%%%%%%%%%%%%%%%%%%%%%%%%%%%%%%%%%%%%%%%%%%%%%%%%%%%%%%%%%%%%%
%%%%%%%%%%%%%%%%%%%%%%%%%%%%%%%%%%%%%%%%%%%%%%%%%%%%%%%%%%%%%%%%%%%%%%%%%%%%%%
\section{Introduction}\label{intro}
\setcounter{equation}{0}

Spin structure of the proton has been studied extensively.
Although it is not still obvious how to interpret the proton spin
in terms of partons, it is important to test our knowledge of spin
physics by using other observables.
The spin structure of spin-1 hadrons is a good example. In particular,
tensor structure exists for the spin-1 hadrons as a new
ingredient. However, it is not clear at this stage how to
describe it in a parton model although there are some initial 
studies \cite{fs,mit-b1,ck-b1,spin-1}.
In this sense, the studies of spin-1 hadrons should be
a challenging experience. 

On the other hand, the Relativistic Heavy Ion Collider (RHIC)
will be completed soon and new proton-proton (pp) polarization
experiments will be done. As a next-generation project, 
the spin-1 deuteron acceleration may be possible \cite{rhic-d}.
With this future project and others \cite{rcnp} in mind,
we have completed general formalism for the polarized Drell-Yan
processes with spin-1/2 and spin-1 hadrons in Ref. \cite{our1}.
It was revealed that 108 structure functions exist in the reactions
and that the number becomes 22 after integrating the cross section
over the virtual-photon transverse momentum $\vec Q_T$ or after taking
the limit $Q_T\rightarrow 0$. 

The purpose of this paper is to clarify how these structure functions
are related to the parton distributions in the colliding hadrons.
In a parton model, it is known that the unpolarized distribution
$f_1$ (or denoted as $q$), the longitudinally polarized one $g_1$
($\Delta q$), and the transversity one $h_1$ ($\Delta_T q$) are studied
in the pp Drell-Yan processes. In addition to these, the tensor distribution
$b_1$ ($\delta q$) should contribute to our Drell-Yan processes.
In the following sections, we obtain the cross section in the parton model
and then discuss how $b_1$ is related to one of the
structure functions in Ref. \cite{our1}. We also study how spin
asymmetries can be expressed by the parton distributions.

First, the hadron tensor is obtained in the parton model by finding
possible Lorentz vector and pseudovector combinations in section \ref{pd-dy}.
Then, the cross section is calculated by using the hadron tensor.
Second, we derive the cross-section expression
by integrating over $\vec Q_T$ or by taking the limit $Q_T\rightarrow 0$
in section \ref{qt}. Third, using the $\vec Q_T$-integrated
cross section, we express the spin asymmetries in terms of the parton
distributions in \ref{asymm}. Finally, our studies are summarized
in section \ref{sum}.

%%%%%%%%%%%%%%%%%%%%%%%%%%%%%%%%%%%%%%%%%%%%%%%%%%%%%%%%%%%%%%%%%%%%%%%%%%%%%%
%%%%%%%%%%%%%%%%%%%%%%%%%%%%%%%%%%%%%%%%%%%%%%%%%%%%%%%%%%%%%%%%%%%%%%%%%%%%%%
\section{Parton-model description of the Drell-Yan process}
\label{pd-dy}
\setcounter{equation}{0}

%%%%%%%%%%%%%%%%%%%%%%%%%%%%%%%%%%%%%%%%%%%%%%%%%%%%%%%%%%%%%%%%%%%%%%%%%%%%%%
\subsection{Drell-Yan process}

Parton-model analyses of the Drell-Yan process with spin-1/2 hadrons
were reported by Ralston-Soper \cite{rs}, Donohue-Gottlieb \cite{dg},
and Tangerman-Mulders (TM) \cite{tm}. In order to clarify the difference
from the spin-1/2 case, we discuss our Drell-Yan process along
the TM formalism. We consider the following process
\begin{equation}
A \, (spin \, 1/2) + B \, (spin \, 1) \rightarrow \ell^+ \ell^- +X
\ ,
\end{equation}
where $A$ and $B$ are spin-1/2 and spin-1 hadrons, respectively.
They could be any hadrons; however, realistic ones are
the proton and the deuteron experimentally. The description
of this paper can be applied to any hadrons with spin-1/2 and spin-1.

Although the formalism of this subsection is discussed in Ref. \cite{tm},
we explain it in order to present the definitions of kinematical variables
and functions for understanding the subsequent sections.
The cross section is written in terms of the lepton tensor
$L_{\mu\nu}$ and the hadron tensor $W^{\mu\nu}$ as \cite{our1}
\begin{equation}
\frac{d\sigma}{d^4 Q \, d \Omega} = \frac{\alpha^2}{2 \, s \, Q^4} 
                                    \, L_{\mu\nu} \, W^{\mu\nu}
\ ,
\label{eqn:cross-0}
\end{equation}
where $\alpha=e^2/(4\pi)$ is the fine structure constant,
$s$ is the center-of-mass energy squared $s=(P_A+P_B)^2$,
$Q$ is the total dilepton momentum, and $\Omega$ is the solid
angle of the momentum $\vec k_{\ell^+}-\vec k_{\ell^-}$.
The hadron and lepton masses are neglected in comparison with
$s$ and $Q^2$: $M_A^2, M_B^2 \ll s$ and $m_\ell^2 \ll Q^2$.
In order to compare with the TM results, we use the same notations
as many as we can. Three vectors $\hat x^\mu$, $\hat y^\mu$,
and $\hat z^\mu$ are defined as 
\begin{equation}
\hat x^\mu = - \frac{X^\mu}{\sqrt{-X^2}} \ , \ \ \ 
\hat y^\mu = - \frac{Y^\mu}{\sqrt{-Y^2}} \ , \ \ \ 
\hat z^\mu = + \frac{Z^\mu}{\sqrt{-Z^2}} \ , 
\end{equation}
where $X^\mu$, $Y^\mu$, and $Z^\mu$ are given in Ref. \cite{our1}.
The $Z^\mu$ axis corresponds to the Collins-Soper choice \cite{cs}
and these axes are spacelike in the dilepton rest frame.
In the same way, $\hat Q$ is defined by
\begin{equation}
\hat Q^\mu = \frac{Q^\mu}{\sqrt{Q^2}}
\ .
\end{equation}
It is convenient to introduce the lightcone notation
$a = [a^-, a^+, \vec a_T]$, where $a^\pm = (a^0 \pm a^3)/\sqrt{2}$.
The transverse vector $a_T =[0,0,\vec a_T]$ is projected out by
\begin{equation}
g_T^{\mu\nu} = g^{\mu\nu} - n_+^\mu n_-^\nu - n_+^\nu n_-^\mu 
\ ,
\end{equation}
with $n_+ = [0,\kappa,\vec 0_T]$ and $n_- = [1/\kappa,0,\vec 0_T]$,
where $\kappa=\sqrt{x_A/x_B}$ in the center-of-momentum (c.m.) frame. 
The transverse vector is orthogonal to the hadron momenta:
$a_T\cdot P_A=a_T\cdot P_B=0$. 
Furthermore, we define $g_\perp^{\mu\nu}$ as
\begin{equation}
g_\perp^{\mu\nu} = g^{\mu\nu} - \hat Q^\mu \hat Q^\nu
                              + \hat z^\mu \hat z^\nu 
\ ,
\end{equation}
which projects out the perpendicular vector
$a_\perp^\mu \equiv g_\perp^{\mu\nu} a_{T \nu}$. It is orthogonal to
the vectors $\hat Q$ and $\hat z$:
$a_\perp \cdot \hat Q = a_\perp \cdot \hat z = 0$.
Because the transverse projection is equal to the perpendicular one
if the $1/Q$ term can be neglected
[$g_T^{\mu\nu}=g_\perp^{\mu\nu}+O(1/Q)$], we use the approximation
$a_\perp \approx a_T$ in the following calculations.
Nevertheless, the perpendicular vectors are often used because
they are convenient due to the orthogonal relation.
The lepton tensor is expressed in terms of these quantities as
\cite{tm}
\begin{align}
L^{\mu \nu} = & \, 2 \, k_{\ell^+}^\mu \, k_{\ell^-}^\nu 
               +2 \, k_{\ell^+}^\nu \, k_{\ell^-}^\mu - Q^2 \, g^{\mu\nu}
\nonumber \\
=& - \frac{Q^2}{2} \, \bigg[ \, (1 + \cos^2 \theta) \, g_\perp^{\mu \nu}  
     - 2 \sin^2 \theta \, \hat{z}^\mu \, \hat{z}^\nu
     +2 \sin^2 \theta \, \cos 2 \phi \,
          (\hat{x}^\mu \hat{x}^\nu + \frac{1}{2} \, g_\perp^{\mu \nu}) 
\nonumber \\
   & \ \ \ \ \ \ \ \ \ 
     + \sin^2 \theta \, \sin 2 \phi \, \hat{x}^{\{ \mu} \hat{y}^{\nu \}}
     + \sin 2 \theta \, \cos   \phi \, \hat{z}^{\{ \mu} \hat{x}^{\nu \}}
     + \sin 2 \theta \, \sin   \phi \, \hat{z}^{\{ \mu} \hat{y}^{\nu \}}
   \, \bigg ]
\ ,
\label{eqn:lepton}
\end{align}
where $\theta$ and $\phi$ are polar and azimuthal angles of the 
vector $\vec k_{\ell^+} - \vec k_{\ell^-}$, and the notation
$A^{ \{ \mu} B^{\nu \} }$ is defined by
\begin{equation}
A^{ \{ \mu} B^{\nu \} } \equiv A^\mu B^\nu + A^\nu B^\mu
\ .
\label{eqn:brace}
\end{equation}

The hadron tensor is given by
\begin{equation}
W^{\mu \nu} = \int \frac{d^4\xi}{(2\pi)^4} \, e^{iQ \cdot \xi} \, 
   < P_A S_A ; P_B S_B \, | \, J^\mu (0) \, J^\nu (\xi) 
                            \, | \, P_A S_A ; P_B S_B >
\ ,
\label{eqn:w-1}
\end{equation}
where $J^\mu$ is the electromagnetic current, and
the hadron momenta and spins are denoted as $P_A$, $P_B$, $S_A$,
and $S_B$. The analysis of the hadron tensor is more complicated
than the one in deep inelastic lepton-hadron scattering because
it contains the currents with two-hadron states. 
The leading lightcone singularity originates from
the process that a quark emits a virtual photon,
which then splits into $\ell^+\ell^-$. However, it does not contribute
to the cross section significantly because the quark should be far
off-shell \cite{rlj}. The dominant contribution comes from
quark-antiquark annihilation processes.
In the following, we discuss the hadron tensor and the cross section
due to the annihilation process: 
$q$(in A)+$\bar q$(in B)$\rightarrow \ell^+ + \ell^-$
in Fig. \ref{fig:qqbar}. Of course, the opposite process
$\bar q$(in A)+$q$(in B)$\rightarrow \ell^+ + \ell^-$ should be
taken into account in order to compare with the experimental cross
section. Its contribution is included in discussing
the spin asymmetries in section \ref{asymm}.
The first process contribution to the hadron tensor in Eq. (\ref{eqn:w-1})
is
\begin{equation}
W^{\mu \nu} = \frac{1}{3} \sum_{a, b} \delta_{b \bar{a}} \, e_a^2 
            \int d^4 k_a \, d^4 k_b \, \delta^4 (k_a + k_b - Q) \, 
            Tr [\Phi_{a/A} (P_A S_A; k_a) \gamma^\mu
           \bar{\Phi}_{b/B} (P_B S_B; k_b) \gamma^\nu]
\ ,
\label{eqn:w-2}
\end{equation}
where $k_a$ and $k_b=k_{\bar a}$ are the quark and antiquark momenta,
the color average is taken by the factor $1/3=3\cdot (1/3)^2$,
and $e_a$ is the charge of a quark with the flavor $a$.
The correlation functions $\Phi_{a/A}$ and $\bar \Phi_{\bar a/B}$ are
defined by \cite{rs}
\begin{align}
\Phi_{a/A} (P_A S_A; k_a)_{ij} & \equiv 
           \int \frac{d^4 \xi}{(2 \pi)^4} \, e^{i k_a \cdot \xi}
<P_A S_A \, | \, \bar{\psi}_j^{(a)}(0) \, \psi_i^{(a)}(\xi) \, | \, P_A S_A>
\ ,
\nonumber \\
\bar{\Phi}_{\bar a/B} (P_B S_B; k_{\bar a})_{ij} & \equiv 
           \int \frac{d^4 \xi}{(2 \pi)^4} \, e^{i k_{\bar a} \cdot \xi}
<P_B S_B \, | \, \psi_i^{(a)}(0) \, \bar{\psi}_j^{(a)}(\xi) \, | \, P_B S_B>
\ .
\end{align}
Link operators should be introduced in these matrix elements so as to
become gauge invariant \cite{tm} although they are not explicitly
written in the above equations. It is also known that they
become identity in the lightcone gauge. In any case, such link operators
do not alter the following discussions of this paper
in a naive parton model.
Figure \ref{fig:qqbar} suggests that the hadron tensor could be written
by a product of quark-hadron amplitudes. 
However, the matrix indices in the trace
of Eq. (\ref{eqn:w-2}) look like 
$[\bar \psi_j^{(a)} (0) \,      \psi_i^{(a)} (\xi)]_A \,
 [\bar \psi_\ell^{(a)} (\xi) \, \psi_k^{(a)} (0)]_B \,
 (\gamma^\mu)_{jk} \, (\gamma^\nu)_{\ell i}$, which is not in a separable
form. A Fierz transformation \cite{ps-book} is used so that the index
summations are taken separately in the hadrons $A$ and $B$.
Using the relation
\begin{align}
\! \! \! \! \! \! \! \! 
4 (\gamma^\mu)_{jk} (\gamma^\nu)_{li} = & \, [{\bf 1}_{ji} {\bf1}_{lk}
           + (i \gamma_5)_{ji} (i \gamma_5)_{lk} 
           - (\gamma^\alpha)_{ji} (\gamma_\alpha)_{lk}
           - (\gamma^\alpha \gamma_5)_{ji} 
                      (\gamma_\alpha \gamma_5)_{lk} 
           + \frac{1}{2} (i \sigma_{\alpha \beta} \gamma_5)_{ji}
                      (i \sigma^{\alpha \beta} \gamma_5)_{lk}] \, g^{\mu \nu}
\nonumber \\
       &   + (\gamma^{\{ \mu})_{ji} (\gamma^{\nu \}})_{lk}
           + (\gamma^{\{ \mu} \gamma_5)_{ji}
                      (\gamma^{\nu \}} \gamma_5)_{lk}
           + (i \sigma^{\alpha \{ \mu} \gamma_5)_{ji}
                      (i \sigma^{\nu \}}_{\ \ \ \alpha} \gamma_5)_{lk}
\ ,
\label{eqn:fierz}
\end{align}
we factorize the hadron tensor as
\begin{align}
& W^{\mu \nu} = \frac{1}{3} \sum_{a, b} \delta_{b \bar{a}} \, e_a^2 
      \int d^2 \vec{k}_{aT} \, d^2 \vec{k}_{bT} \,
      \delta^2 (\vec{k}_{aT} + \vec{k}_{bT} - \vec{Q}_T) \, 
      \bigg[ \, \bigg\{ \, 
    - \Phi_{a/A} [\gamma^\alpha] \, \bar{\Phi}_{b/B}[\gamma_\alpha]
\nonumber \\
&   - \Phi_{a/A} [\gamma^\alpha \gamma_5] \,
      \bar{\Phi}_{b/B}[\gamma_\alpha \gamma_5] 
    + \frac{1}{2} \Phi_{a/A} [i \sigma_{\alpha \beta} \gamma_5] \,
      \bar{\Phi}_{b/B}[i \sigma^{\alpha \beta} \gamma_5] \bigg\} \, g^{\mu\nu} 
    + \Phi_{a/A} [\gamma^{\{\mu}] \, \bar{\Phi}_{b/B}[\gamma^{\nu\}}]
\nonumber \\
&   + \Phi_{a/A} [\gamma^{\{\mu}\gamma_5] \,
      \bar{\Phi}_{b/B}[\gamma^{\nu\}}\gamma_5]
    + \Phi_{a/A} [i \sigma^{\alpha\{\mu}\gamma_5] \, 
      \bar{\Phi}_{b/B}[i \sigma_{\ \ \, \alpha}^{\nu\}}\gamma_5] \, \bigg] \, 
    + O(1/Q)
\ .
\label{eqn:w-3}
\end{align}
The definition of the brace is given in Eq. (\ref{eqn:brace}).
For example, the last term in the bracket is explicitly written as
$\Phi_{a/A} [i \sigma^{\alpha\{\mu}\gamma_5] \, 
      \bar{\Phi}_{b/B}[i \sigma_{\ \ \, \alpha}^{\nu\}}\gamma_5]
= \Phi_{a/A} [i \sigma^{\alpha \mu}\gamma_5] \, 
      \bar{\Phi}_{b/B}[i \sigma_{\ \, \alpha}^{\nu}\gamma_5]
+ \Phi_{a/A} [i \sigma^{\alpha \nu}\gamma_5] \, 
      \bar{\Phi}_{b/B}[i \sigma_{\ \, \alpha}^{\mu}\gamma_5]$.
The functions $\Phi_{a/A} [\Gamma]$ and $\bar \Phi_{b/B} [\Gamma]$ are
defined by
\begin{align}
\Phi_{a/A} [\Gamma] (x, \vec{k}_T) & \equiv  
              \frac{1}{2} \int dk_a^- \, Tr [\Gamma \, \Phi_{a/A}]
\ ,
\label{eqn:phi-a}
\\
\bar{\Phi}_{b/B} [\Gamma] (x, \vec{k}_T) & \equiv  
              \frac{1}{2} \int dk_b^+ \, Tr [\Gamma \, \bar{\Phi}_{b/B}]
\ ,
\end{align}
and we assume $k_b^+ \ll k_a^+$ and $k_a^- \ll k_b^-$
in obtaining the factorized expression.
The functions $\Phi_{a/A}[{\bf 1}]$ and $\Phi_{a/A}[i\gamma_5]$
are obtained as $\Phi_{a/A}[{\bf 1}]\sim O(1/Q)$ and 
$\Phi_{a/A}[i\gamma_5]=0$ according to
the calculations in the next subsection, so that they are
not explicitly written in Eq. (\ref{eqn:w-3}).
As it is obvious from Eq. (\ref{eqn:w-3}), we do not address ourselves to
the higher-twist terms. Because the Drell-Yan process of spin-1/2 and
spin-1 hadrons is not investigated at all in any parton model,
we first discuss the leading contributions in this paper.
We found 108 (22) structure functions in general (after integration
over $\vec Q_T$), and most of them are related to the higher-twist
physics as it becomes obvious in the later sections of this paper.
Although it is interesting to study the higher-twist terms \cite{htwist},
we leave this topic as our future project.

%%%%%%%%%%%%%%%%%%%%%%%%%%%%%%%%%%%%%%%%%%%%%%%%%%%%%%%%%%%%%%%%%%%%%%%%%%%%%%
\subsection{Correlation functions and parton distributions}
\label{phi}

The hadron tensor is expressed by the correlation functions
in Eq. (\ref{eqn:w-2}). We expand them in terms
of the possible Lorentz vectors and pseudovectors.
The correlation function $\Phi (P S; k)$ is a matrix with
sixteen components, so that it can be written in terms of
sixteen $4\times 4$ matrices \cite{ps-book}:
\begin{equation}
{\bf 1},\   \gamma_5,\   \gamma^\mu,\   \gamma^\mu \gamma_5,\ 
\sigma^{\mu \nu} \gamma_5  
\ .
\label{eqn:matrices}
\end{equation}
Because the spin-1/2 hadron case was already discussed in Refs.
\cite{rs,dg,tm}, we investigate the correlation function for a
spin-1 hadron. We discuss it in the frame where the spin-1 hadron
is moving in the $z$ direction. Because the $z_{cm}$ direction is
taken as the direction of the hadron $A$ momentum in Ref. \cite{our1},
it could mean that we assume the hadron $A$ as if it were
a spin-1 hadron in this subsection. Of course, appropriate
spin-1/2 and spin-1 expressions are used for the hadron $A$ and $B$,
respectively, in calculating the hadron tensor and the cross section.
The correlation function is expanded in terms of the matrices
of Eq. (\ref{eqn:matrices}) together with the possible Lorentz
vectors and pseudovector: $P^\mu$, $k^\mu$, and $S^\mu$.
However, these combinations have to satisfy the usual conditions
of Hermiticity, parity conservation, and time-reversal invariance.
In finding the possible combinations, we should be careful
that the rank-two spin terms are allowed for a spin-1 hadron.
The reader may look at Ref. \cite{our1} for the detailed discussions
on this point. Considering these conditions, we obtain the possible
Lorentz scalar quantities. Then, the coefficient $A_i$ is assigned
for each term:
\begin{align}
\Phi (P & S; k ) =  A_1 {\bf 1} + A_2 \psl 
                      + A_3 \ksl 
                      + A_4 \gamma_5 \ssl
                      + A_5 \gamma_5 [\psl, \ssl] 
                      + A_6 \gamma_5 [\ksl, \ssl]
                     + A_7 k \cdot S \gamma_5 \psl
                     + A_8 k \cdot S \gamma_5 \ksl
\nonumber \\ &
                     + A_9 k \cdot S \gamma_5 [\psl, \ksl]
                     + A_{10} (k \cdot S)^2 {\bf 1} 
                     + A_{11} (k \cdot S)^2 \psl
                     + A_{12} (k \cdot S)^2 \ksl
                     + A_{13} \, k \cdot S \, \ssl
\ .
\label{eqn:phi-1}
\end{align}
For simplicity, the subscripts $a$ and $A$ in $\Phi$, momenta,
momentum fraction, spin, and helicity are not written in this
subsection.
The spin dependent factors $(k \cdot S)^2$ and $k \cdot S \, \ssl$
do not exist in a spin-1/2 hadron,
so that the terms $A_{10}$, $A_{11}$, $A_{12}$, and $A_{13}$ are
the additional ones to the spin-1/2 expression in Ref. \cite{tm}.
It means that the interesting tensor structure of the spin-1 hadron
is contained in these new terms.

With the general expression of Eq. (\ref{eqn:phi-1}),
we can calculate $\Phi[\Gamma]$ in Eq. (\ref{eqn:w-3}).
Because the new terms contribute only to $\Phi[\gamma^\alpha]$,
its calculation procedure is discussed in the following
by using the lightcone representation.
The $\gamma$ matrices are given by \cite{rlj}
\begin{equation}
\gamma^0 =
   \begin{pmatrix} 
              0          & \sigma^3           \\
          \sigma^3       & 0                  \\
   \end{pmatrix}
\ , \ \ \ 
\vec \gamma_{_T} =
   \begin{pmatrix} 
   i \, \vec\sigma_{_T}  & 0                    \\
              0          & i \, \vec\sigma_{_T} \\
   \end{pmatrix}
\ , \ \ \ 
\gamma^3 =
   \begin{pmatrix} 
              0          & -\sigma^3          \\
         +\sigma^3       & 0                  \\
   \end{pmatrix}
\ , \ \ \ 
\gamma^\pm = \frac{1}{\sqrt{2}} \, (\gamma^0 \pm \gamma^3 )
\ .
\end{equation}
It is necessary to calculate $\Phi[\gamma^+]$, $\Phi[\gamma^-]$, 
and $\Phi[\vec \gamma_{_T}]$ for obtaining the $\Phi [\gamma^\alpha]$
term in Eq. (\ref{eqn:w-3}).
Substituting Eq. (\ref{eqn:phi-1}) into Eq. (\ref{eqn:phi-a}), we
have the expression for $\Phi [\gamma^+]$ as
\begin{align}
\Phi [\gamma^+] & = \frac{1}{2} \int dk^- Tr [\gamma^+ \Phi]
\nonumber \\
                & = \frac{1}{2} \int dk^- Tr \left[ \, \gamma^+
                  \{ A_2 \psl + A_3 \ksl + A_{11} (k\cdot S)^2 \psl
              + A_{12} (k\cdot S)^2 \ksl+ A_{13} \, k\cdot S \, \ssl \} 
                      \, \right ]
\ .
\end{align}
The traces of the lightcone $\gamma$ matrices are
\begin{alignat}{2}
& Tr(\gamma^+ \gamma^+)=Tr(\gamma^- \gamma^-)=0, 
  \ \ \ \ & &
  Tr(\gamma^+ \gamma^-)=Tr(\gamma^- \gamma^+)=4,
\nonumber \\
& Tr(\gamma^+ \vec \gamma_{_T})=Tr(\gamma^- \vec \gamma_{_T})=0,
  \ \ \ \ & &
  Tr(\gamma_{_T}^i \gamma_{_T}^j)=-4 \, \delta_T^{ij},
\ .
\end{alignat}
Therefore, the trace $Tr(\gamma^+ \psl)$ becomes $Tr(\gamma^+ \psl)=4 P^+$.
Then, introducing the momentum fraction $x$ and the helicity $\lambda$
by $k^+=xP^+$ and $S^+=\lambda P^+/M$, we obtain
\begin{equation}
\Phi [\gamma^+] = \int d(2k\cdot P) \bigg [ A_2 + x A_3 
               + (\vec k_T \cdot \vec S_T)^2 \, (A_{11} + x A_{12})
               -  \vec k_T \cdot \vec S_T \, \frac{\lambda}{M} \, A_{13}
                  \bigg ]
\ ,
\end{equation}
where $M$ is the hadron mass. This equation is written as
\begin{equation}
\Phi [\gamma^+] = f_1(x, \vec{k}_T^{\,\, 2})
           + b_1(x, \vec{k}_T^{\,\, 2}) 
        \left[\frac{4 (\vec{k}_T \cdot \vec{S}_T)^2}{\vec{k}_T^{\,\, 2}}
            - \frac{2}{3} \right] 
            + c_1(x, \vec{k}_T^{\,\, 2}) 
              \, \lambda \, \frac{\vec{k}_T \cdot \vec{S}_T}{M} 
\ , 
\end{equation}
with the parton distributions
\begin{align}
f_1 (x, \vec{k}_T^{\,\, 2}) = &\int d (2 k \cdot P) \,  
         \left\{ A_2 + \vec{k}_T^{\,\, 2}  A_{11}/6 
         +x \left(A_3 + \vec{k}_T^{\,\, 2} A_{12} / 6 \right) \right\}
\ , \nonumber \\
b_1 (x, \vec{k}_T^{\,\, 2}) = &\int d (2 k \cdot P) \, 
         \vec{k}_T^{\,\, 2} \, (A_{11}+ x A_{12}) / 4
\ , \nonumber \\
c_1 (x, \vec{k}_T^{\,\, 2}) = &  - \int d (2 k \cdot P) \, A_{13}
\ .
\end{align}
Because the integration variable $2k\cdot P$ is constrained by
the relation $\vec{k}_T^{\,\, 2} + k^2 - 2 x k \cdot P + x^2 M^2 = 0$,
it is of the order of $1$ despite $P^+ \sim O(Q)$.
In addition to the usual unpolarized parton distribution $f_1$,
there appear new distributions $b_1$ and $c_1$ which do not exist
for a spin-1/2 hadron. 
The correlation function $\Phi[\gamma^+]$ can be also expressed by
the quark probability density ${\mathcal P}(x,\vec k_T)$
according to Ref. \cite{tm}:
\begin{equation}
\Phi [\gamma^+] = {\mathcal P} (x,\vec k_T)
\ .
\end{equation}
Integrating over the transverse momentum $\vec k_T$, we obtain
\begin{equation}
{\mathcal P}(x) = f_1(x) + b_1(x) \, \frac{2}{3} \, 
                              (2 \, |\vec S_T|^2 -\lambda^2)
\ ,
\label{eqn:unpol-1}
\end{equation}
where the relation $\lambda^2+|\vec S_T|^2=1$ is used and the functions
${\mathcal P}(x)$, $f_1(x)$, and $b_1(x)$ are defined by
\begin{align}
{\mathcal P}(x) & = \int d^2 \vec k_T \, {\mathcal P}(x,\vec k_T)
\ , \nonumber \\
f_1 (x) & = \int d^2 \vec k_T \, f_1 (x,\vec k_T^2)
\ , \nonumber \\
b_1 (x) & = \int d^2 \vec k_T \, b_1 (x,\vec k_T^2)
\ .
\end{align}
It is obvious from the spin combination in Eq. (\ref{eqn:unpol-1})
that $b_1$ is related to the tensor structure.
Equation (\ref{eqn:unpol-1}) means that the $b_1$ distribution 
is given by
\begin{equation}
b_1(x)= \frac{1}{2} \, \bigg[ \, {\mathcal P}(x)_{\lambda=0} 
          -  \frac{ {\mathcal P}(x)_{\lambda=+1} 
                   +{\mathcal P}(x)_{\lambda=-1} }{2} \, \bigg]
\ .
\label{eqn:b1x-p}
\end{equation}
Indeed, this definition of the tensor distribution agrees with the one 
in the lepton-deuteron studies \cite{mit-b1}.
On the other hand, $c_1$ should be related to the intermediate
polarization according to Ref. \cite{our1}. This distribution has
never been discussed as far as we are aware, so that it is simply
named $c_1$. This is a brand-new one in this paper. 
The interesting point of this distribution is that it cannot be
measured by the longitudinally and transversely polarized reactions.
The optimum way of observing it is to polarize the spin-1 hadron
with the angles $45^\circ$ and $135^\circ$ with
respect to the hadron momentum direction. However,
as we mention in section \ref{asymm}, it is difficult to attain
the intermediate polarization in the collider experiments
if the deuteron is used as a spin-1 hadron because of its small
magnetic moment.
It is not unique to express $c_1$ in terms of ${\mathcal P}(x,\vec k_T)$.
For example, it is written by the distributions in the intermediate
polarizations $I_1$ and $I_3$ of Ref. \cite{our1} as
\begin{equation}
{\mathcal P}(x,\vec k_T)_{I_1} - {\mathcal P}(x,\vec k_T)_{I_3} 
  = \frac{| \vec k_T |}{M} \, \sin \phi_k \, c_1(x, \vec{k}_T^{\,\, 2}) 
\ ,
\end{equation}
where $\phi_k$ is the azimuthal angle of the vector $\vec k_T$.
Because its contribution vanishes by integrating the correlation
function over $\vec Q_T$ or by taking the limit $Q_T\rightarrow 0$,
it could be related to higher-twist distributions. This point should
be clarified by our future project. At this stage, we should content
ourselves with the studies of leading contributions because there
exists no parton-model analysis of the Drell-Yan process with a spin-1
hadron. The same calculations are done for the functions $\Phi [\gamma^-]$
and $\Phi [\gamma_{_T}]$. However, these are proportional to $O(1/Q)$,
so that they are ignored in the following discussions.

The other correlation functions are calculated in the same
way; however, the results are the same as those in Ref. \cite{tm}:
\begin{align}
\Phi [\gamma^+ \gamma_5] & = {\mathcal P} (x,\vec k_T) \,
                                 \lambda (x,\vec k_T) 
              = g_{1L}(x, \vec{k}_T^{\,\, 2}) \, \lambda
     + g_{1T}(x, \vec{k}_T^{\,\, 2}) \, \frac{\vec{k}_T \cdot \vec{S}_T}{M}
\ , \nonumber \\
\Phi [i \sigma^{i +} \gamma_5] & = {\mathcal P} (x,\vec k_T) \,
                                   \vec s_{T}^{\,\, i} (x,\vec k_T) 
 = h_{1T}(x, \vec{k}_T^{\,\, 2}) \, \vec{S}_T^{\,\, i} 
               + \left[h_{1L}^\perp(x, \vec{k}_T^{\,\, 2}) \, \lambda 
               + h_{1T}^\perp(x, \vec{k}_T^{\,\, 2}) \, 
               \frac{\vec{k}_T \cdot \vec{S}_T}{M}
               \right] \frac{\vec{k}_T^{\,\, i}}{M} 
\ .
\label{eqn:gh}
\end{align}
Here, $\lambda (x,\vec k_T)$ and $\vec s_{T}^{\,\, i} (x,\vec k_T)$
are the quark helicity and transverse polarization densities.
The longitudinally and transversely polarized distributions
are given as
\begin{alignat}{2}
g_{1 L} (x, \vec{k}_T^{\,\, 2}) & = \int d (2 k \cdot P) \, (A_4 / M)
\ , \ \ \ \ & 
g_{1 T} (x, \vec{k}_T^{\,\, 2}) & = - \int d (2 k \cdot P) \, 
         M(A_7 + x A_8)
\ , \nonumber \\
h_{1 T} (x, \vec{k}_T^{\,\, 2}) & = - \int d (2 k \cdot P) \, 
         2 \, (A_5 + x A_6)
\ , \ \ \ \ & 
h_{1 L}^\perp (x, \vec{k}_T^{\,\, 2}) & = \int d (2 k \cdot P) \, 2 \, A_6
\ , \nonumber \\
h_{1 T}^\perp (x, \vec{k}_T^{\,\, 2}) & = \int d (2 k \cdot P) \, 
         2 \, M^2 A_9 
\ .
& &
\end{alignat}
Because the explanations were given for the $g_1$ and $h_1$
distributions in Ref. \cite{tm}, we do not repeat them in this paper.
The $A_{10}$ term contributes to $\Phi[{\bf 1}]$ as an additional
one to the spin-1/2 case; however, it is proportional to $O(1/Q)$.
The terms with $\Phi[{\bf 1}]$ and $\Phi[i \gamma_5]$ are excluded from
our formalism because the relations $\Phi[{\bf 1}]\sim O(1/Q)$
and $\Phi[i \gamma_5]=0$ are obtained by the similar calculations.

In calculating the Drell-Yan cross section, the antiquark
correlation function should be also expressed in terms of
the antiquark distributions.
It is obtained by using the charge-conjugation property \cite{tm,bd}.
The antiquark correlation function in the hadron $A$ is related to 
the quark one in the antihadron $\bar A$ by
the charge-conjugation matrix $C=i \gamma^2 \gamma^0$:
\begin{equation}
\bar{\Phi}_{\bar{a} / A} 
     = - \, C^{-1} \, \left( \Phi_{a / \bar{A}} \right)^T \, C
\ ,
\end{equation}  
where the superscript $T$ indicates the transposed matrix.
Because the quark distribution in the antihadron is equal
to the antiquark distribution in the hadron,
the antiquark correlation functions is expressed as
\begin{align}
\bar{\Phi} [\gamma^+] =& \bar{f}_1(x, \vec{k}_T^{\,\, 2}) 
            + \bar b_1(x, \vec{k}_T^{\,\, 2}) 
         \left[\frac{4 (\vec{k}_T \cdot \vec{S}_T)^2}{\vec{k}_T^{\,\, 2}}
            - \frac{2}{3} \right] 
            + \bar c_1(x, \vec{k}_T^{\,\, 2}) \,  
              \lambda \, \frac{\vec{k}_T \cdot \vec{S}_T}{M}  
\ , \nonumber \\
\bar{\Phi} [\gamma^+ \gamma_5] =& 
              -\bar{g}_{1L}(x, \vec{k}_T^{\,\, 2}) \, \lambda
              - \bar{g}_{1T}(x, \vec{k}_T^{\,\, 2}) \,
              \frac{\vec{k}_T \cdot \vec{S}_T}{M} 
\ , \nonumber \\
\bar{\Phi} [i \sigma^{i +} \gamma_5] =& \bar{h}_{1T}(x, \vec{k}_T^{\,\, 2}) 
              \, \vec{S}_T^i 
             + \left[\bar{h}_{1L}^\perp(x, \vec{k}_T^{\,\, 2}) \, \lambda 
             + \bar{h}_{1T}^\perp(x, \vec{k}_T^{\,\, 2}) \,
             \frac{\vec{k}_T \cdot \vec{S}_T}{M}
            \right] \frac{\vec{k}_T^{\,\, i}}{M}
\ .
\end{align}
Furthermore, the anticommutation relations for fermions indicate
$\bar \Phi_{ij} (PS; k) = - \Phi_{ij} (PS; - k)$, so that the
distributions satisfy the relation
\begin{equation}
\bar f (x, \vec k_T^{\,\, 2}) =
          \left\{ 
          \begin{array}{ll}
          - f (-x, \vec k_T^{\,\, 2})  \ \ \ \ 
                   & \textrm{$f=f_1$, $b_1$, $g_{1T}$,
                                          $h_{1T}$, and $h_{1T}^\perp$} \\
          + f (-x, \vec k_T^{\,\, 2}) \ \ \ \ 
                   &  \textrm{$f=c_1$, $g_{1L}$,
                                           and $h_{1L}^\perp$} \ .  \\
         \end{array} 
         \right. 
\end{equation}                                                

In this way, we have derived the expressions of the correlation
functions in terms of the quark and antiquark distributions.
As new distributions, the $b_1$, $\bar b_1$, $c_1$, and $\bar c_1$
distributions appear in the correlation functions.
In particular, the $b_1$ distribution is a leading-twist one
and it is associated with the tensor structure of the spin-1 hadron.
On the other hand, the $c_1$ distribution is related to the intermediate polarization of Ref. \cite{our1} and it could be associated with
higher-twist physics. The other longitudinally and transversely polarized
distributions exist in the same way as those of a spin-1/2 hadron.

%%%%%%%%%%%%%%%%%%%%%%%%%%%%%%%%%%%%%%%%%%%%%%%%%%%%%%%%%%%%%%%%%%%%%%%%%%%%%%
\subsection{Structure functions and the cross section}
\label{structure}

Because the correlation functions in Eq. (\ref{eqn:w-3}) are calculated,
the hadron tensor can be expressed in terms of the parton distributions.
Neglecting the higher-twist contributions, we write the hadron
tensor as
\begin{align}
& W^{\mu \nu} = -\frac{1}{3} \sum_{a, b} \delta_{b \bar{a}} \, e_a^2 
      \int d^2 \vec{k}_{aT} \, d^2 \vec{k}_{bT} \,
      \delta^2 (\vec{k}_{aT} + \vec{k}_{bT} - \vec{Q}_T) \,
      \bigg\{ \left( \, \Phi_{a/A} [\gamma^+] \, \bar{\Phi}_{b/B}[\gamma^-] 
         \right.
\nonumber \\
  &  \left.
     + \Phi_{a/A} [\gamma^+ \gamma_5] 
        \, \bar{\Phi}_{b/B}[\gamma^- \gamma_5] \, \right) \, g^{\mu \nu}_T  
     + \Phi_{a/A} [i \sigma^{i+} \gamma_5] \, 
         \bar{\Phi}_{b/B}[i \sigma^{j-} \gamma_5] \,  
         \left(g_{Ti}^{\ \ \{ \mu} g_{T\ \ \ j}^{\ \ \nu \}}
                 - g_{Tij} \, g^{\mu \nu}_T \right) \biggr\} 
\ .
\label{eqn:w-4}
\end{align}
The correlation functions in the previous subsection are substituted
into the above equation. Then, the integrals over $\vec k_{aT}$
and $\vec k_{aT}$ are manipulated by using the equations in
Appendix of Ref. \cite{tm}. The calculations are rather lengthy
particularly in the transverse $h_1$ part. However,
because the $g_1$ and $h_1$ portions are the same as the ones
in a spin-1/2 hadron  \cite{tm} and the calculations of $f_1$, $b_1$,
and $c_1$ terms are rather simple, we do not explain the calculation
procedure. Noting that the $b_1$ and $c_1$ terms do not exist for
the spin-1/2 hadron $A$, we obtain
\begin{align}
& W^{\mu \nu} =  - g_\perp^{\mu \nu} \, \bigg [ \, 
                   W_T 
                    + \frac{1}{4} \, \lambda_A \, \lambda_B \, V_T^{LL} 
                    - \frac{2}{3} \, V_T^{UQ_0(1)} 
                    - 2 \, S_{B\perp}^2 \, V_T^{UQ_0(2)}
                    + 2 \, (\hat x \cdot  S_{B\perp})^2 \, V_T^{UQ_0(3)}
\nonumber \\
&                - \lambda_B \, \hat x \cdot S_{B\perp} \, V_T^{UQ_1}
                 - \lambda_A \, \hat x \cdot S_{B\perp} \, V_T^{LT}
                 - \hat x \cdot S_{A\perp} \, \lambda_B \, V_T^{TL}
                 - S_{A\perp} \cdot S_{B\perp} \, V_T^{TT(1)}
                + \hat x \cdot S_{A\perp} \,
                   \hat x \cdot S_{B\perp} \, V_T^{TT(2)} \, \bigg ]
\nonumber \\
& - \left( \hat{x}^\mu \hat{x}^\nu + \frac{1}{2} \, g_\perp^{\mu \nu} \right)
       \, \bigg[ \frac{1}{4} \, \lambda_A \, \lambda_B \, V_{2,2}^{LL} 
                 - \lambda_A \, \hat x \cdot S_{B\perp} \, V_{2,2}^{LT}
           - \hat x \cdot S_{A\perp} \, \lambda_B \, V_{2,2}^{TL}
           - S_{A\perp} \cdot S_{B\perp} \, V_{2,2}^{TT(1)}
\nonumber \\
&            + \hat x \cdot S_{A\perp} \,
                   \hat x \cdot S_{B\perp} \, V_{2,2}^{TT(2)} \, \bigg ]
   + \left ( \hat{x}^{\{ \mu} S_{A \perp}^{\nu \}} 
            - \hat{x} \cdot S_{A \perp} \, g_\perp^{\mu \nu} \right )
       \left ( \hat{x} \cdot S_{B \perp} \, U_{2,2}^{TT(A)}
              -\lambda_B \, U_{2,2}^{TL} \right )  
\nonumber \\
&  + \left ( \hat{x}^{\{ \mu} S_{B \perp}^{\nu \}} 
            - \hat{x} \cdot S_{B \perp} \, g_\perp^{\mu \nu} \right )
       \left ( \hat{x} \cdot S_{A \perp} \, U_{2,2}^{TT(B)}
              -\lambda_A \, U_{2,2}^{LT} \right )  
 - \left ( S_{A \perp}^{\{ \mu} S_{B \perp}^{\nu \}} 
           - S_{A \perp} \cdot S_{B \perp} \, g_\perp^{\mu \nu} \right )
             U_{2, 2}^{TT}
\ .
\label{eqn:w-5}
\end{align}
The structure functions are expressed by the integral
\begin{equation}
I[d_1 \bar{d}_2] \equiv \frac{1}{3} 
      \sum_{a, b} \delta_{b \bar{a}} \, e_a^2 
      \int d^2 \vec{k}_{aT} \, d^2 \vec{k}_{bT} \,
      \delta^2 (\vec{k}_{aT} + \vec{k}_{bT} - \vec{Q}_T) 
       \, d_1 (x_A, \vec{k}_{aT}^{\,\, 2}) \,
          \bar{d}_2 (x_B, \vec{k}_{bT}^{\,\, 2})
\ .
\label{eqn:i}
\end{equation}
First, the unpolarized structure function is given by
\begin{equation}
W_T  = I[f_1 \bar{f}_1]
\ ,
\label{eqn:wt}
\end{equation}
where the subscript $T$ of $W_T$ corresponds to the index combination of
$(0,0)-(2,0)/3$ in the expressions of Ref. \cite{our1}.
The structure functions associated with the factors $-g_\perp^{\mu\nu}$
and $\hat z^\mu \hat z^\nu$ in $W^{\mu\nu}$ are denoted as
$W_T$ and $W_L$. Obviously, $W_L$ vanishes in the parton model. 
The longitudinal structure functions $V_T^{LL}$ and $V_{2,2}^{LL}$ are
\begin{align}
V_T^{LL} =& - 4 \, I[g_{1L} \, \bar{g}_{1L}]
\ , \nonumber \\
V_{2, 2}^{LL} =& \,
     I \left[ \left( \alpha + \beta - \frac{(\alpha - \beta)^2}{Q_T^2}\right)
          \frac{4 \, h_{1L}^\perp \, \bar{h}_{1L}^\perp}{M_A M_B}\right] 
\ ,
\end{align}
where $Q_T^2$ is given by $\vec Q_T^2$ (note $Q_T^2 \ne -\vec Q_T^2$) and
the variables $\alpha$ and $\beta$ are defined 
by $\alpha = \vec{k}_{aT}^{\,\, 2}$ and $\beta = \vec{k}_{bT}^{\,\, 2}$.
The tensor structure functions become
\begin{align}
V_T^{U Q_0 (1)} =& \, I[f_1 \bar{b}_1]
\ , \nonumber \\
V_T^{U Q_0 (2)} =& \, I \left[ \left(-Q_T^2 + 2(\alpha + \beta) 
          - \frac{(\alpha - \beta)^2}{Q_T^2} \right)
          \frac{f_1 \, \bar{b}_1}{2 \, \beta} \right]
\ , \nonumber \\
V_T^{U Q_0 (3)} =& \, I \left[ \left(Q_T^2 - 2 \alpha 
          + \frac{(\alpha - \beta)^2}{Q_T^2} \right)
            \frac{f_1 \, \bar{b}_1}{\beta} \right]    
\ , \nonumber \\
V_T^{U Q_1} =& \, I \left[(Q_T^2 - \alpha + \beta) 
          \frac{f_1 \, \bar{c}_1}{2 \, M_B \, Q_T} \right] 
\ .
\label{eqn:sf-uq}
\end{align}
The superscripts $U$, $Q_0$, and $Q_1$ indicate
the unpolarized state $U$, quadrupole polarization $Q_0$,
and quadrupole polarization $Q_1$, respectively \cite{our1}.
For example, $V_T^{U Q_1}$ indicates that the hadron $A$ is
unpolarized and $B$ is polarized with the quadrupole
polarization $Q_1$.
The longitudinal-transverse structure functions are
\begin{align}
V_T^{LT} =& \, I \left[ ( -Q_T^2 +\alpha - \beta) 
          \frac{ g_{1L} \, \bar{g}_{1T}}{2 \, M_B \, Q_T}\right] 
\ , \nonumber \\
V_T^{TL} =& \, I \left[ ( -Q_T^2 -\alpha + \beta) 
          \frac{ g_{1T} \, \bar{g}_{1L}}{2 \, M_A \, Q_T}\right] 
\ , \nonumber \\
V_{2, 2}^{LT} =& \,
     I \left[ \left(\alpha \, Q_T^2 +\beta^2 + \alpha \beta -2 \alpha^2
          + \frac{(\alpha - \beta)^3}{Q_T^2}\right) 
      \frac{h_{1L}^\perp \, \bar{h}_{1T}^\perp}{M_A \, M_B^2 \, Q_T}\right] 
\ , \nonumber \\
V_{2, 2}^{TL} =& \,
     I \left[ \left(\beta \, Q_T^2 +\alpha^2 + \alpha \beta -2 \beta^2
          - \frac{(\alpha - \beta)^3}{Q_T^2}\right) 
      \frac{h_{1T}^\perp \, \bar{h}_{1L}^\perp}{M_A^2 \, M_B \, Q_T}\right] 
\ , \nonumber \\
U_{2, 2}^{LT} =& \,
     I \left[(Q_T^2 +\alpha - \beta)
          \frac{h_{1L}^\perp \, \bar{h}_{1T}}{2 M_A Q_T} 
    - \left(Q_T^2(\alpha - \beta) -2 (\alpha^2 - \beta^2) 
          + \frac{(\alpha - \beta)^3}{Q_T^2}\right) 
       \frac{h_{1L}^\perp \, \bar{h}_{1T}^\perp}{4 \, M_A \, M_B^2 \, Q_T}
            \right] 
\ , \nonumber \\
U_{2, 2}^{TL} =& \,
     I \left[(Q_T^2 -\alpha + \beta)
          \frac{h_{1T} \, \bar{h}_{1L}^\perp}{2 \, M_B \, Q_T} 
    + \left(Q_T^2(\alpha - \beta) -2 (\alpha^2 - \beta^2) 
          + \frac{(\alpha - \beta)^3}{Q_T^2}\right) 
       \frac{h_{1T}^\perp \, \bar{h}_{1L}^\perp}{4 \, M_A^2 \, M_B \, Q_T}
          \right] 
\ .
\label{eqn:sf-parton}
\end{align}
In the Drell-Yan process of identical hadrons, the structure functions
$V^{TL}$ and $U^{TL}$ are equal to $V^{LT}$ and $U^{LT}$.
However, they are different in our reactions, so that both types are listed. 
The transverse structure functions are
\begin{align}
V_T^{TT (1)} =& \,
     I \left[ \left( -Q_T^2 +2 \alpha + 2 \beta
          - \frac{(\alpha - \beta)^2}{Q_T^2}\right) 
          \frac{ g_{1T} \, \bar{g}_{1T}}{4 \, M_A \, M_B}\right] 
\ , \nonumber \\
V_T^{TT (2)} =& \,
     I \left[ \left(-\alpha - \beta
          + \frac{(\alpha - \beta)^2}{Q_T^2}\right) 
          \frac{ g_{1T} \, \bar{g}_{1T}}{2 \, M_A \, M_B}\right] 
\ , \nonumber \\
V_{2, 2}^{TT (1)} =& \,
    I \left[ \left( Q_T^2 (\alpha+\beta) - (\alpha-\beta)^2 
                     - 2 (\alpha+\beta)^2 
          + \frac{3 (\alpha + \beta) (\alpha - \beta)^2}{Q_T^2}
          - \frac{(\alpha - \beta)^4}{Q_T^4}\right) 
   \frac{h_{1T}^\perp \, \bar{h}_{1T}^\perp}{4 \, M_A^2 \, M_B^2 }\right] 
\ , \nonumber \\
V_{2, 2}^{TT (2)} =& \,
     I \left[ \left(\alpha^2 + \beta^2 
          - \frac{2 (\alpha + \beta) (\alpha - \beta)^2}{Q_T^2}
          + \frac{(\alpha - \beta)^4}{Q_T^4}\right) 
          \frac{h_{1T}^\perp \, \bar{h}_{1T}^\perp}{M_A^2 \, M_B^2 }\right] 
\ , \nonumber \\
U_{2,2}^{TT(A)} = & \, I \bigg[
                   \left ( Q_T^2 - 2 \alpha 
                   + \frac{(\alpha-\beta)^2}{Q_T^2} \right ) 
                  \frac{h_{1T} \bar{h}_{1T}^\perp}{2 \, M_B^2} 
                + \bigg ( (\alpha-\beta) Q_T^2 + (\alpha-\beta)^2 
                          - 4 \alpha (\alpha-\beta) 
\nonumber \\
               & \ \ \ \ \ \ \ \ \ \ \ \ \ 
                   +\frac{ 2\alpha(\alpha-\beta)^2
                          +(\alpha-\beta)^2 (\alpha+\beta) }{Q_T^2}
                   - \frac{(\alpha-\beta)^4}{Q_T^4}  \bigg ) 
     \frac{h_{1T}^\perp \, \bar{h}_{1T}^\perp}{8 \, M_A^2 \, M_B^2} \bigg]
\ ,
\nonumber \\
U_{2,2}^{TT(B)} = & \, I \bigg[
                   \left ( Q_T^2 - 2 \beta 
                   + \frac{(\alpha-\beta)^2}{Q_T^2} \right ) 
                  \frac{h_{1T}^\perp \bar{h}_{1T}}{2 \, M_A^2} 
                + \bigg ( - (\alpha-\beta) Q_T^2 + (\alpha-\beta)^2 
                          +4 \beta (\alpha-\beta) 
\nonumber \\
               & \ \ \ \ \ \ \ \ \ \ \ \ \ 
                   +\frac{ 2\beta(\alpha-\beta)^2
                          +(\alpha-\beta)^2 (\alpha+\beta) }{Q_T^2}
                   - \frac{(\alpha-\beta)^4}{Q_T^4}  \bigg ) 
   \frac{h_{1T}^\perp \, \bar{h}_{1T}^\perp}{8 \, M_A^2 \, M_B^2} \bigg]
\ , \nonumber \\
U_{2,2}^{TT} = & \, I  \bigg [ h_{1T} \bar{h}_{1T} 
                 - \left ( Q_T^2 - 2 \alpha - 2 \beta
                   + \frac{(\alpha-\beta)^2}{Q_T^2} \right ) 
                 \left( 
                    \frac{h_{1T}^\perp \bar{h}_{1T}}{4 \, M_A^2}
                  + \frac{h_{1T} \bar{h}_{1T}^\perp}{4 \, M_B^2} \right) 
                        \bigg ]
\ .
\label{eqn:vutt}
\end{align}

Substituting the hadron tensor of Eq. (\ref{eqn:w-5}) and 
the lepton tensor of Eq. (\ref{eqn:lepton}) into Eq. (\ref{eqn:cross-0}),
we obtain the cross section
\begin{align}
& \frac{d \sigma}{d^4Q \, d \Omega} = \frac{\alpha^2}{2 \, s \, Q^2} \,
      \bigg [ \,  (1 + \cos^2 \theta) \bigg\{ \, 
                W_T + \frac{1}{4} \lambda_A \lambda_B \, V_T^{LL}
               - \frac{2}{3} \, V_T^{UQ_0(1)} 
               + 2 \, |\vec S_{BT}|^2 \, V_T^{UQ_0(2)}
\nonumber \\
& \ \ \ \ 
               + 2 \, |\vec S_{BT}|^2 \cos^2 \phi_B \, V_T^{UQ_0(3)}
               + \lambda_B \, |\vec S_{BT}| \cos \phi_B \, V_T^{UQ_1}
               + \lambda_A  \, |\vec S_{BT}| \cos \phi_B \, V_T^{LT} 
\nonumber \\
& \ \ \ \ 
               + \lambda_B \, |\vec S_{AT}| \cos \phi_A \, V_T^{TL} 
  + |\vec S_{AT}| \, |\vec S_{BT}| \cos(\phi_A-\phi_B) \, V_T^{TT(1)} 
  + |\vec S_{AT}| \, |\vec S_{BT}| 
                  \cos \phi_A \cos \phi_B \, V_T^{TT(2)} \, \bigg\}
\nonumber \\
& \ \ \ \ 
   + \sin^2 \theta \, \bigg\{ \, \frac{1}{2} \cos 2\phi \, \bigg ( \,
       \frac{1}{4} \lambda_A \lambda_B \, V_{2,2}^{LL}
     +             \lambda_A \, |\vec S_{BT}| \cos \phi_B \, V_{2,2}^{LT}
     +             \lambda_B \, |\vec S_{AT}| \cos \phi_A \, V_{2,2}^{TL}
\nonumber \\
& \ \ \ \ 
     +   |\vec S_{AT}| \, |\vec S_{BT}| \cos(\phi_A-\phi_B) 
                   (V_{2,2}^{TT(1)}+U_{2,2}^{TT(A)}+U_{2,2}^{TT(B)})
     +   |\vec S_{AT}| \, |\vec S_{BT}|
                   \cos\phi_A \cos\phi_B \, V_{2,2}^{TT(2)} \, \bigg )
\nonumber \\
& \ \ \ \ 
  + |\vec S_{AT}| \cos(2\phi-\phi_A) \, \lambda_B \, U_{2,2}^{TL} 
  + |\vec S_{BT}| \cos(2\phi-\phi_B) \, \lambda_A \, U_{2,2}^{LT} 
\nonumber \\
& \ \ \ \ 
  + \frac{1}{2} \sin 2\phi \,
              |\vec S_{AT}| \, |\vec S_{BT}| \,  \sin(\phi_A-\phi_B) \,
                       (U_{2,2}^{TT(A)}-U_{2,2}^{TT(B)})
\nonumber \\
& \ \ \ \ 
  + |\vec S_{AT}| \, |\vec S_{BT}| \cos(2\phi-\phi_A-\phi_B) 
            (U_{2,2}^{TT} + U_{2,2}^{TT(A)}/2 + U_{2,2}^{TT(B)}/2) 
        \, \bigg\} \, \bigg ]
\ .
\label{eqn:cross-1}
\end{align}
This equation indicates that $V_{2,2}^{TT(1)}$, $U_{2,2}^{TT(A)}$,
$U_{2,2}^{TT(B)}$, and $U_{2,2}^{TT}$ cannot be measured independently.
Only the combinations $V_{2,2}^{TT(1)}+U_{2,2}^{TT(A)}+U_{2,2}^{TT(B)}$,
$U_{2,2}^{TT(A)}-U_{2,2}^{TT(B)}$, and
$U_{2,2}^{TT} + U_{2,2}^{TT(A)}/2 + U_{2,2}^{TT(B)}/2$ can be
studied experimentally. 
In our previous paper \cite{our1}, we predicted that 108 structure
functions exist in the Drell-Yan processes of spin-1/2 and spin-1
hadrons. According to Eq. (\ref{eqn:cross-1}), there are 19 independent
ones in the naive parton model.
It means that the rest of them are related to the neglected
$O(1/Q)$ terms, namely the higher-twist structure functions. 
Although the $V_T^{UQ_0(1)}$ term may seem to contribute to
the unpolarized cross section, it is canceled out by the other
$UQ_0$-type terms in taking the spin average.

%%%%%%%%%%%%%%%%%%%%%%%%%%%%%%%%%%%%%%%%%%%%%%%%%%%%%%%%%%%%%%%%%%%%%%%%%%%%%%
%%%%%%%%%%%%%%%%%%%%%%%%%%%%%%%%%%%%%%%%%%%%%%%%%%%%%%%%%%%%%%%%%%%%%%%%%%%%%%
\section{$\bf \vec Q_T$ integration and the limit $\bf Q_T\rightarrow 0$}
\label{qt}
\setcounter{equation}{0}

Because the 108 structure functions are too many to investigate seriously
and many of them are not important at this stage,
the cross section is integrated over $\vec Q_T$ or it
is calculated in the limit $Q_T\rightarrow 0$ \cite{our1}.
There exist 22 structure functions in these cases, and they are
considered to be physically significant. On the other hand,
the cross section of Eq. (\ref{eqn:cross-1}) is obtained at
finite $Q_T$. In order to compare with the results in Ref .\cite{our1},
we should investigate the parton-model cross section
by taking the $\vec Q_T$ integration or the limit $Q_T\rightarrow 0$.

%%%%%%%%%%%%%%%%%%%%%%%%%%%%%%%%%%%%%%%%%%%%%%%%%%%%%%%%%%%%%%%%%%%%%%%%%%%%%%
\subsection{$\bf \vec Q_T$-integrated cross section}
\label{qt-int}

First, we discuss the integration of the cross section over $\vec Q_T$.
If the hadron tensor Eq. ({\ref{eqn:w-4}) is integrated over $\vec Q_T$, 
the delta function $\delta^2 (\vec{k}_{aT} + \vec{k}_{bT} - \vec{Q}_T)$
disappears. It means that the integrations over $\vec k_{aT}$ and 
$\vec k_{bT}$ can be calculated separately. 
Then, the integrals with odd functions of $\vec k_T$ vanish: e.g. 
$\int d^2 \vec k_{aT} F (\vec k_{aT}^{\,\, 2}, \vec k_{bT}^{\,\, 2})
     \vec k_{aT}=0$.
Furthermore, the vector $Q_T^\mu$ can not be used any longer in
expanding, for example, the integral 
$\int d^2 \vec k_{aT} d^2 \vec k_{bT} 
  F (\vec k_{aT}^{\,\, 2}, \vec k_{bT}^{\,\, 2}) 
  k_{1\perp}^\mu k_{2\perp}^\nu$
in terms of the possible Lorentz-vector combinations.
We calculate the hadron tensor in the similar way with the one
in section \ref{structure}. However, the calculations are much simpler
because we do not have to take into account $Q_T^\mu$.
It does not make sense to explain the calculation procedure again,
so that only the final results are shown in the following. 

The $\vec Q_T$-integrated hadron tensor is expressed as
\begin{equation}
\overline W^{\, \mu\nu} = \int d^2 \vec Q_T \, W^{\mu\nu}
\ ,
\end{equation}
and in the same way for the structure functions.
Then, the hadron tensor becomes
\begin{align}
\overline W^{\, \mu\nu} = & - g_\perp^{\mu\nu} \, \bigg\{ 
     \overline W_T  
    + \frac{1}{4} \, \lambda_A \lambda_B \, \overline V_T^{\, LL}
    - 2 \left( S_{B\perp}^{\, 2} +\frac{1}{3} \right) \, 
                     \overline V_T^{\, UQ_0} \, \bigg\}
\nonumber \\
&  -\left ( S_{A \perp}^{\{ \mu} S_{B \perp}^{\nu \}} 
           - S_{A \perp} \cdot S_{B \perp} \, g_\perp^{\mu \nu} \right )
      \overline U_{2,2}^{\, TT} 
\ ,
\label{eqn:w-int}
\end{align}
where the structure functions are written in terms of the 
parton distributions as
\begin{align}
\overline W_T     & = \frac{1}{3} \sum_a e_a^2 \, 
                              f_1(x_A) \, \bar f_1(x_B)
\ , \nonumber \\
\overline V_T^{\, LL} & = - \frac{4}{3} \sum_a e_a^2 \, 
                              g_1(x_A) \, \bar g_1(x_B)
\ , \nonumber \\
\overline U_{2,2}^{\, TT} & = \frac{1}{3} \sum_a e_a^2 \, 
                              h_1(x_A) \, \bar h_1(x_B)
\ , \nonumber \\
\overline V_T^{\, UQ_0} & \equiv \overline V_T^{\, UQ_0(1)}
                                = \overline V_T^{\, UQ_0(2)}
            = \frac{1}{3} \sum_a e_a^2 \, 
                              f_1(x_A) \, \bar b_1(x_B)
\ .
\label{eqn:sf-int}
\end{align}
The quark distributions are defined in the $\vec k_T$-integrated form as
\begin{align}
f (x) & = \int d^2 \vec{k}_T \, f(x, \vec{k}_T^{\,\, 2}) \, \ \ \ \ 
\textrm{for $f$=$f_1$, $g_1(=g_{1L})$, and $b_1$}
\ , \\
h_1 (x) & = \int d^2 \vec k_T \left[ h_{1T}(x, \vec{k}_T^{\,\, 2}) 
          + \frac{\vec{k}_T^{\,\, 2}}{2 M^2} 
               h_{1T}^\perp(x, \vec{k}_T^{\,\, 2}) \right]
\ ,
\end{align}
and in the same way for the antiquark distributions.
The other structure functions vanish by the $\vec Q_T$ integration:
\begin{align}
& \overline V_T^{\, UQ_0(3)} = 
\overline V_T^{\, UQ_1} = 
\overline V_T^{\, LT} = 
\overline V_T^{\, TL} = 
\overline V_T^{\, TT(1)} = 
\overline V_T^{\, TT(2)} = 
\overline V_{2,2}^{\, LL} = 
\overline V_{2,2}^{\, LT} 
\nonumber \\
& = \overline V_{2,2}^{\, TL} = 
\overline V_{2,2}^{\, TT(1)} = 
\overline V_{2,2}^{\, TT(2)} = 
\overline U_{2,2}^{\, LT} = 
\overline U_{2,2}^{\, TL} = 
\overline U_{2,2}^{\, TT(A)} = 
\overline U_{2,2}^{\, TT(B)} = 0
\ .
\end{align}
With the expression of the hadron tensor in Eq. (\ref{eqn:w-int}),
the cross section becomes
\begin{align}
\frac{d \sigma}{dx_A \, dx_B \, d \Omega} = \frac{\alpha^2}{4 \, Q^2} \,
      \bigg [ \,  & (1 + \cos^2 \theta) \bigg\{ \, 
            \overline W_T 
            + \frac{1}{4} \lambda_A \lambda_B \, \overline V_T^{\, LL}
            + \frac{2}{3}  \left( 2 \, |\vec S_{BT}|^2 - \lambda_B^2 \right)
                               \, \overline V_T^{\, UQ_0} \, \bigg\}
\nonumber \\
&  +  \sin^2 \theta \, |\vec S_{AT}| \, |\vec S_{BT}|
             \cos(2\phi-\phi_A-\phi_B) \, \overline U_{2,2}^{\, TT}
        \, \bigg ]
\ .
\label{eqn:cross-int}
\end{align}
The tensor distribution $b_1$ contributes to the cross section through
the structure function $\overline V_T^{\, UQ_0}$. 
Because it is given by the multiplication of $f_1$ and $\bar b_1$
($\bar f_1$ and $b_1$ in the opposite process) in Eq. (\ref{eqn:sf-int}),
the quark and antiquark tensor distributions could be measured 
if the unpolarized distributions in the hadron $A$
are well known. The $\bar b_1$ is paired with $f_1$; however, it is not
with $g_{1L}$ and $h_1$. This is because of the 
Fierz transformation in Eq. (\ref{eqn:w-3}): 
$\Phi_{a/A}[\gamma^+]$ is multiplied by $\bar \Phi_{b/B}[\gamma^-]$
and not by the other factors $\bar \Phi_{b/B}[\gamma^- \gamma_5]$
and $\bar{\Phi}_{b/B}[i \sigma^{j-} \gamma_5]$,
and the tensor distributions appear only in the functions
$\Phi_{a/A} [\gamma^+]$ and $\bar \Phi_{b/B} [\gamma^-]$.
Therefore, $\bar b_1$ can couple only with $f_1$, $b_1$, and $c_1$.
Since the hadron $A$ is a spin-1/2 particle, the distributions $b_1$
and $c_1$ do not exist. In this way, the only possible combination
is $f_1(x_A) \bar b_1(x_B)$ for the process
$q$(in A)+$\bar q$(in B)$\rightarrow \ell^+ + \ell^-$.

%%%%%%%%%%%%%%%%%%%%%%%%%%%%%%%%%%%%%%%%%%%%%%%%%%%%%%%%%%%%%%%%%%%%%%%%%%%%%%
\subsection{Cross section in the limit $\bf Q_T\rightarrow 0$}
\label{qt=0}

The transverse momentum $Q_T$ is generally small in comparison with
the dilepton mass $Q$. It originates mainly from intrinsic transverse momenta
of the partons, so that its magnitude is roughly restricted by
the hadron size $r$: $Q_T \lesssim 1/r$. In this respect, it makes sense
to consider the limit $Q_T\rightarrow 0$ for finding the essential part. 

The structure functions in Eqs. (\ref{eqn:wt})$-$(\ref{eqn:vutt})
should be evaluated in this limit.
As an example, we show how to take the limit for $V_T^{UQ_0(2)}$
in Eq. (\ref{eqn:sf-uq}). 
The integration variables $\vec k_{aT}$ and $\vec k_{bT}$
in Eq. (\ref{eqn:i}) are changed for $\vec k_T$ and $\vec K_T$,
which are defined by
$\vec k_T = (\vec k_{aT} - \vec k_{bT})/2$ and
$\vec K_T = \vec k_{aT} + \vec k_{bT}$.
Then, the delta function is integrated out and the structure
function is expressed as 
\begin{align}
V_T^{UQ_0(2)} & = \frac{1}{3} \sum_a e_a^2 
                   \, \frac{1}{2 \, (\vec k_T - \vec Q_T /2)^2}
                              \left\{ 4 \, \vec k_T^{\,\, 2} -
             \frac{4 \, (\vec k_T\cdot \vec Q_T)^2}{Q_T^2} \right\} 
\nonumber \\ 
& \ \ \ \ \ \ \ \ \ \times
               f_1 \left( x_A,(\vec k_T+\vec Q_T/2)^2 \right) \, 
             \bar b_1 \left( x_B,(\vec k_T-\vec Q_T/2)^2 \right)
\nonumber \\
         & = I_0 [f_1 \bar b_1] \ \ \ \ \textrm{in $Q_T\rightarrow 0$}
\ ,
\label{eqn:vuq0}
\end{align}
where the function $I_0$ is defined by
\begin{equation}
I_0 [d_1 \bar{d}_2] = \frac{1}{3}  \sum_a e_a^2 \int d^2 \vec{k}_{T}  
         \,  d_1 (x_A, \vec{k}_{T}^{\,\, 2}) \, 
             \bar{d}_2 (x_B, \vec{k}_{T}^{\,\, 2})
\ ,
\end{equation}
and $k_T^i k_T^j/\vec k_T^{\,\, 2}$ in the second term of Eq. (\ref{eqn:vuq0})
is replaced by $\delta_T^{ij}/2$. 
The other structure functions are calculated in the same way.
The finite ones are obtained as
\begin{align}
& W_T = I_0 [f_1 \bar f_1] \ , \ \ \ 
  V_T^{LL}  = -4 I_0 [g_{1L} \, \bar g_{1L}] \ , \ \ \ 
  V_T^{UQ_0(1)} = I_0 [f_1 \bar b_1] 
                = V_T^{UQ_0(2)} \equiv V_T^{UQ_0} 
\ , \nonumber \\
& V_T^{\, TT(1)} =  I_0 \left[ \frac{\vec k_T^{\,\, 2}}{2 \, M_A \, M_B} \,
                        g_{1T} \, \bar g_{1T} \right ]
\ , \ \ \ 
U_{2,2}^{TT} = I_0 \left[ h_{1T} \, \bar h_{1T} 
                + \frac{\vec k_T^{\,\, 2}}{2} \left ( 
                            \frac{h_{1T}^\perp \, \bar h_{1T}}{M_A^2}
                          + \frac{h_{1T} \, \bar h_{1T}^\perp}{M_B^2} \right )
                    \, \right ]
\ , \nonumber \\
& U_{2,2}^{TT(A)}  = I_0 \left[ \frac{\vec k_T^{\,\, 4}}{4 \, M_A^2 \, M_B^2} 
                       \,  h_{1T}^\perp \, \bar h_{1T}^\perp \right ]
                 = U_{2,2}^{TT(B)}
                 = - \frac{1}{2} V_{2,2}^{TT(1)}
\ .
\end{align}   
The following ones vanish in the $Q_T\rightarrow 0$ limit:
\begin{align}
& V_T^{\, UQ_0(3)} = 
  V_T^{\, UQ_1} = 
  V_T^{\, LT} = 
  V_T^{\, TL} = 
  V_T^{\, TT(2)} =
  V_{2,2}^{\, LL}
\nonumber \\ 
& = 
  V_{2,2}^{\, LT} =
  V_{2,2}^{\, TL} = 
  V_{2,2}^{\, TT(2)} = 
  U_{2,2}^{\, LT} = 
  U_{2,2}^{\, TL} = 0
\ .
\end{align}
The hadron tensor is expressed by the finite structure functions as
\begin{align}
W^{\, \mu\nu} =  & - g_\perp^{\mu\nu} \, \bigg\{ W_T  
    + \frac{1}{4} \, \lambda_A \lambda_B \, V_T^{\, LL}
    - 2 \left( S_{B\perp}^2 +\frac{1}{3} \right) \, V_T^{\, UQ_0} 
    - S_{A\perp} \cdot S_{B\perp} \, V_T^{TT(1)} \, \bigg\}
\nonumber \\
& 
+ \left ( \hat{x}^\mu \hat{x}^\nu + \frac{1}{2} \, g_\perp^{\mu \nu} \right )
        S_{A\perp} \cdot S_{B\perp} \, V_{2,2}^{TT(1)}
   + \left ( \hat{x}^{\{ \mu} S_{A \perp}^{\nu \}} 
            - \hat{x} \cdot S_{A \perp} \, g_\perp^{\mu \nu} \right )
       \hat{x} \cdot S_{B \perp} \, U_{2,2}^{TT(A)}
\nonumber \\
& 
+ \left ( \hat{x}^{\{ \mu} S_{B \perp}^{\nu \}} 
            - \hat{x} \cdot S_{B \perp} \, g_\perp^{\mu \nu} \right )
       \hat{x} \cdot S_{A \perp} \, U_{2,2}^{TT(B)}
  -\left ( S_{A \perp}^{\{ \mu} S_{B \perp}^{\nu \}} 
           - S_{A \perp} \cdot S_{B \perp} \, g_\perp^{\mu \nu} \right )
      U_{2,2}^{TT} 
\ .
\end{align}
Defining $\hat U_{2, 2}^{TT}$ by
\begin{align}
\hat U_{2, 2}^{TT} & = U_{2, 2}^{TT} + U_{2, 2}^{TT(A)}/2 
                                     + U_{2, 2}^{TT(B)}/2
\nonumber \\
&   = I_0 [h_1 \bar h_1] 
\ ,
\end{align}
we obtain the cross section as
\begin{align}
\frac{d \sigma}{d^4 Q \, d \Omega} & = 
     \frac{\alpha^2}{2 \, s \, Q^2}  \,
     \bigg[ \, (1 + \cos^2 \theta) \bigg\{ W_T
             + \frac{1}{4} \, \lambda_A \lambda_B \, V_T^{LL} 
     + \frac{2}{3} \, (2 |\vec{S}_{BT}|^2 - \lambda_B^2) \, V_T^{U Q_0} 
\nonumber \\   
    & + |\vec S_{AT}| \, |\vec S_{BT}| \cos(\phi_A-\phi_B) \, V_T^{TT(1)} 
\bigg\}
   + \sin^2 \theta \, |\vec{S}_{AT}| \, |\vec{S}_{BT}| 
              \cos (2 \phi - \phi_A - \phi_B) \,
              \hat U_{2, 2}^{TT} \, \bigg]    
\ . 
\label{eqn:cross-qt0}
\end{align} 
Even though $V_{2,2}^{TT(1)}$ is finite, it does not contribute
to the cross section. It is canceled out by the term 
$U_{2,2}^{TT(A)}+U_{2,2}^{TT(B)}$.
As it is obvious from the results of this subsection and
section \ref{qt-int}, the expressions of the hadron tensors
and the cross sections are slightly different in the $\vec Q_T$ 
integration and in the $Q_T\rightarrow 0$ limit.

Among many structure functions in Eq. (\ref{eqn:cross-1}), we have
extracted the essential ones by taking the limit $Q_T\rightarrow 0$
or by the $\vec Q_T$ integration. According to the expressions
of the cross section in Eqs. (\ref{eqn:cross-int}) and
(\ref{eqn:cross-qt0}), merely the four or five structure functions
remain finite: the unpolarized structure function $W_T$, the longitudinal
one $V_T^{LL}$, the transverse one(s) $U_{2,2}^{TT}$ ($V_T^{TT(1)}$), 
and the unpolarized-quadrupole one $V_T^{U Q_0}$. Most of them are
already known in the pp Drell-Yan reactions. The last quadrupole
structure function $V_T^{U Q_0}$ is new in the Drell-Yan process
of spin-1/2 and spin-1 hadrons.

%%%%%%%%%%%%%%%%%%%%%%%%%%%%%%%%%%%%%%%%%%%%%%%%%%%%%%%%%%%%%%%%%%%%%%%%%%%%%%
%%%%%%%%%%%%%%%%%%%%%%%%%%%%%%%%%%%%%%%%%%%%%%%%%%%%%%%%%%%%%%%%%%%%%%%%%%%%%%
\section{Spin asymmetries and parton distributions}
\label{asymm}
\setcounter{equation}{0}

We have derived the expressions for the Drell-Yan cross section in
the parton model with finite $Q_T$, $\vec Q_T$ integration,
and $Q_T\rightarrow 0$. Because the spin asymmetries are discussed
in Ref. \cite{our1} in the latter two cases, they
are shown in the $\vec Q_T$-integrated case as an example
in this section.
The polarized parton distributions for a spin-1 hadron
are illustrated in Fig. \ref{fig:ltq}.
As it is obvious from Eq. (\ref{eqn:gh}),
the longitudinally polarized (transversity) distribution is defined by
the probability to find a quark with spin polarized along
the longitudinal (transverse) spin of a polarized hadron minus
the probability to find it polarized oppositely.
On the other hand, the tensor polarized distribution $b_1$ is
very different from these distributions according to
Eq. (\ref{eqn:b1x-p}). It is not associated
with the quark polarization as shown by the unpolarized mark $\bullet$ in
Fig. \ref{fig:ltq}. It is related to the ``unpolarized"-quark distribution
in the polarized spin-1 hadron. The tensor distribution is essentially
the difference between the unpolarized-quark distributions in
the longitudinally and transversely polarized hadron states
as it is given in Eq. (\ref{eqn:b1x-p}).

In Ref. \cite{our1}, fifteen spin combinations are suggested . 
However, most of them vanish in the parton model by the $\vec Q_T$
integration. There are only four finite structure functions, and three
of them exist in the pp Drell-Yan processes. 
First, the unpolarized cross section is
\begin{align}
\left < \frac{d\sigma}{dx_A \, dx_B \, d\Omega} \right> & =
         \frac{\alpha^2}{4 \, Q^2} \, (1+\cos^2 \theta) \, \overline W_T
\nonumber \\
  & = \frac{\alpha^2}{4 \, Q^2} \, (1+\cos^2 \theta) 
      \, \frac{1}{3} \sum_a e_a^2 \, 
                  \left [ \, f_1 (x_A) \, \bar f_1 (x_B)
                             + \bar f_1 (x_A) \, f_1 (x_B) \, \right ]
\ .
\end{align}
Next, the longitudinal and transverse double spin asymmetries are
\begin{align}
A_{LL} & = \frac{ \sigma(\downarrow_L , +1_L) 
                       - \sigma(\uparrow_L , +1_L) }
                    { 2 < \! \sigma \! > }
         = -\frac{\overline V_T^{\, LL}}{4 \, \overline W_T}
\nonumber \\
       & = \frac{\sum_a e_a^2 \, 
                \left[ \, g_1(x_A) \, \bar g_1(x_B)
                   + \bar g_1(x_A) \, g_1(x_B) \, \right] }
                {\sum_a e_a^2 \, 
                  \left[ \, f_1 (x_A) \, \bar f_1(x_B)
                          + \bar f_1(x_A) \, f_1(x_B) \, \right] }
\ , \nonumber \\
A_{TT} & = \frac{ \sigma(\phi_A=0 , \phi_B=0) 
                         - \sigma(\phi_A=\pi , \phi_B=0) }
                    { 2 < \! \sigma \! > }
         = \frac{\sin^2\theta \, \cos 2\phi}{1+\cos^2\theta} \,
           \frac{\overline U_{2,2}^{\, TT}}{\overline W_T}
\nonumber \\
         & = \frac{\sin^2\theta \, \cos 2\phi}{1+\cos^2\theta} \,
           \frac{\sum_a e_a^2 \, 
         \left[ \, h_1(x_A) \, \bar h_1(x_B)
                 + \bar h_1(x_A) \, h_1(x_B) \, \right] }
                {\sum_a e_a^2 \, 
                  \left[ \, f_1(x_A) \, \bar f_1(x_B)
                          + \bar f_1(x_A) \, f_1(x_B) \, \right] }
\ ,
\end{align}
where we have explicitly written the contribution from the process
$\bar q$(in A)+$q$(in B)$\rightarrow \ell^+ + \ell^-$
in addition to the one from
$q$(in A)+$\bar q$(in B)$\rightarrow \ell^+ + \ell^-$.
The subscripts of $\uparrow_L$, $\downarrow_L$, and $+1_L$
indicate the longitudinal polarization.
If $\phi_A$ or $\phi_B$ is indicated in the expression
of $\sigma(pol_A,pol_B)$, it means that the hadron $A$ or $B$ is
transversely polarized with the azimuthal angle $\phi_A$ or $\phi_B$.
The above asymmetries are given by the spin-flip cross sections
in the hadron $A$, so that $A_{TT}$ corresponds to $A_{T(T)}$ in
Ref. \cite{our1}. However, the other asymmetry expressions of
Ref. \cite{our1} become the same:
$ A_{TT}^{\, \parallel} = A_{(T)T} = A_{T(T)}$ in the parton model.
Furthermore, the perpendicular transverse-transverse asymmetry
is simply given by 
$A_{TT}^{\, \perp} = \tan 2\phi \, A_{TT}^{\, \parallel} $.
It should be noted that our definitions of the asymmetries are slightly
different from the usual ones in the pp and ep scattering.
The asymmetries are defined so as to exclude the $b_1$
contributions to the denominator. If the usual definition 
$[\sigma(\uparrow -1)-\sigma(\uparrow +1)]/
 [\sigma(\uparrow -1)+\sigma(\uparrow +1)]$ is used,
the $b_1$ contributes, for example, to the longitudinal
asymmetry $A_{LL}$.
The tensor distribution can be investigated by the unpolarized-quadrupole
$Q_0$ asymmetry
\begin{align}
A_{UQ_0} & = \frac{1}{2 < \! \sigma \! >} \,        
         \bigg [ \, \sigma(\bullet , 0_L)
            - \frac{ \sigma(\bullet , +1_L) 
                    +\sigma(\bullet , -1_L) }{2} \, \bigg ]   
         = \frac{\overline V_T^{\, UQ_0}}{\overline W_T}
\nonumber \\
       & = \frac{\sum_a e_a^2 \, 
                  \left[ \, f_1(x_A) \, \bar b_1(x_B)
                          + \bar f_1(x_A) \, b_1(x_B) \, \right] }
                {\sum_a e_a^2 \, 
                  \left[ \, f_1(x_A) \, \bar f_1(x_B)
                          + \bar f_1(x_A) \, f_1(x_B) \, \right] }
\ ,
\end{align}
where the filled circle indicates that the hadron $A$ is unpolarized.
The other asymmetries vanish 
\begin{align}
& A_{LT}=A_{TL}=A_{UT}=A_{TU}=A_{TQ_0}=A_{UQ_1}=A_{LQ_1}=A_{TQ_1}
\nonumber \\
& =A_{UQ_2}=A_{LQ_2}=A_{TQ_2}=0 \ \ \ \ 
   \text{by the $\vec Q_T$ integration.}
\end{align}

The only new finite asymmetry, which does not exist in the pp reactions,
is the unpolarized-quadrupole $Q_0$ asymmetry $A_{UQ_0}$.
In order to measure this quantity, we use a longitudinally and transversely
polarized spin-1 hadron with an unpolarized hadron $A$.
Then, the quadrupole $Q_0$ spin combination \cite{our1} should be taken.
The $b_1$ and $\bar b_1$ distributions could be extracted
from the asymmetry $A_{UQ_0}$ measurements
with the information on the unpolarized parton distributions $f_1(x)$ and
$\bar f_1(x)$ in the hadrons $A$ and $B$.
If it is difficult to attain the longitudinally polarized deuteron
in the collider experiment, we may combine the transversely polarized
cross sections with the unpolarized one without resorting to the
longitudinal polarization. Alternatively, the fixed deuteron target
may be used for obtaining $A_{UQ_0}$.

There is an important advantage to use the Drell-Yan process for measuring
$\bar b_1$ over the lepton scattering. In the large Feynman-$x$
$(x_F=x_A-x_B)$ region, the antiquark distribution $\bar f_1(x_A)$ is
very small in comparison with the quark one $f_1(x_A)$. 
Then, the cross section is dominated by the annihilation process
$q(A)+\bar q(B)\rightarrow \ell^+ + \ell^-$, and the asymmetry becomes
\begin{equation}
A_{UQ_0} \textrm{(large $x_F$)} 
      \approx \frac{\sum_a e_a^2 \, f_1(x_A) \, \bar b_1(x_B)}
                   {\sum_a e_a^2 \, f_1(x_A) \, \bar f_1 (x_B)}
\ .
\end{equation}
This equation means that the antiquark tensor distributions
could be extracted if the unpolarized distributions are well known
in the hadrons $A$ and $B$.
We note in the electron-scattering case
that the $b_1$ sum rule is written by the parton model as
\cite{ck-b1}
\begin{equation}
\int dx \, b_1^e (x)  =  {\displaystyle \lim_{t\rightarrow 0}}
                        - \frac{5}{3} \frac{t}{4M^2} F_Q (t) 
                        + \delta Q_{sea}  
\ ,
\label{eqn:sumb1}
\end{equation}
where $M$ is the hadron mass, $F_Q(t)$ is the quadrupole form factor
in the unit of $1/M^2$, and $\delta Q_{sea}$ is the antiquark tensor
polarization, for example
$\delta Q_{sea}= \int dx  
          [8 \delta \bar u (x) +2 \delta \bar d (x)
           +\delta s (x) +\delta \bar s (x)]/9$ for the deuteron.
The first term of Eq. (\ref{eqn:sumb1}) vanishes, so that the
difference from the sum rule $\int dx \, b_1^e=0$ could suggest
a finite tensor polarization of the antiquarks. 
Although the sum rule is valid within the quark model,
the nuclear shadowing effects should be taken into account
properly \cite{spin-1} as far as the deuteron is concerned.
On the other hand, the antiquark tensor polarization
$\delta \bar q$ (or $\bar b_1$) could be studied independently
by the Drell-Yan process.
As the violation of the Gottfried sum rule leads to the light antiquark
flavor asymmetry $\bar u\ne \bar d$ and the difference was confirmed
by the Drell-Yan experiments \cite{skpr}, the antiquark tensor
polarization could be investigated by both methods: the sum rule
of Eq. (\ref{eqn:sumb1}) in the lepton scattering and the Drell-Yan
process measurement. However, the advantage of the Drell-Yan process
is that the antiquark distribution $\bar b_1(x)$
is directly measured even though it is difficult in the lepton scattering.

Furthermore, the flavor asymmetry in the polarized antiquark distributions
($\Delta \bar u \ne \Delta \bar d$, $\Delta_T \bar u \ne \Delta_T \bar d$)
could be investigated by combining pp and pd Drell-Yan data.
It is particularly important for the transversity distributions
because the $\Delta_T \bar u / \Delta_T \bar d$ asymmetry cannot
be found in the $W^\pm$ production processes due to the chiral-odd
nature.

In this way, we find that a variety of interesting topics are
waiting to be studied in connection with the new structure functions
for spin-1 hadrons. In particular, we have shown in this paper
that the tensor distributions $b_1$ and $\bar b_1$
could be measured by the asymmetry $A_{UQ_0}$ in the Drell-Yan process
of spin-1/2 and spin-1 hadrons. 
A realistic possibility is the proton-deuteron Drell-Yan experiment,
and it may be realized, for example at RHIC \cite{rhic-d}.
In order to support the deuteron polarization project in experimental
high-energy spin physics, we should investigate theoretically more about
the spin structure of spin-1 hadrons.

%%%%%%%%%%%%%%%%%%%%%%%%%%%%%%%%%%%%%%%%%%%%%%%%%%%%%%%%%%%%%%%%%%%%%%%%%%%%%%%%
%%%%%%%%%%%%%%%%%%%%%%%%%%%%%%%%%%%%%%%%%%%%%%%%%%%%%%%%%%%%%%%%%%%%%%%%%%%%%%%%
\section{Summary}\label{sum}

We have investigated the Drell-Yan processes of spin-1/2 and spin-1 hadrons
in a naive parton model by ignoring the $Q(1/Q)$ terms. 
First, the quark and antiquark correlation functions are expressed
by the combinations of possible Lorentz vectors and pseudovectors
by taking into account the Hermiticity, parity conservation, and time-reversal
invariance. Then, we have shown that
the tensor distributions $b_1$ and $c_1$ , which are specific for
a spin-1 hadron, are involved in the correlation function $\Phi [\gamma^+]$.
The expressions of the other functions 
$\Phi [\gamma^+ \gamma_5]$ and $\Phi [i \sigma^{i +} \gamma_5]$
are the same as those of the spin-1/2 hadron in terms of
the longitudinal and transverse distributions.
Using the obtained correlation functions, we have calculated the
hadron tensor and the Drell-Yan cross section.
We found that there exist 19 independent structure functions
in the parton model.
Next, we studied two cases: the $\vec Q_T$ integration and
the limit $Q_T\rightarrow 0$.
In these cases, the $c_1$ contribution vanishes, and
there are only four or five finite structure functions: the unpolarized,
longitudinally polarized, transversely polarized, and 
tensor polarized structure functions. The last one is related to
the tensor polarized distributions $b_1$ and $\bar b_1$.
Although the tensor structure function could be measured in the
lepton scattering, the Drell-Yan measurements
are valuable for finding particularly the antiquark tensor polarization.
In addition to these topics, there are a number of higher-twist structure
functions in the Drell-Yan processes, and they should be also studied
in detail.

%%%%%%%%%%%%%%%%%%%%%%%%%%%%%%%%%%%%%%%%%%%%%%%%%%%%%%%%%%%%%%%%%%%%%%%%%%%%%%
%%%%%%%%%%%%%%%%%%%%%%%%%%%%%%%%%%%%%%%%%%%%%%%%%%%%%%%%%%%%%%%%%%%%%%%%%%%%%%
\section*{{\bf Acknowledgments}}
\addcontentsline{toc}{section}{\protect\numberline{\S}{Acknowledgments}}

S.K. was partly supported by the Grant-in-Aid for Scientific Research
from the Japanese Ministry of Education, Science, and Culture under
the contract number 10640277.
He would like to thank R. L. Jaffe for discussions on spin-1 physics.

%%%%%%%%%%%%%%%%%%%%%%%%%%%%%%%%%%%%%%%%%%%%%%%%%%%%%%%%%%%%%%%%%%%%%%%%%%%%%%
%%%%%%%%%%%%%%%%%%%%%%%%%%%%%%%%%%%%%%%%%%%%%%%%%%%%%%%%%%%%%%%%%%%%%%%%%%%%%%

%%%%%%%%%%%%%%%%%%%%%%%%%%%%%%%%%%%%%%%%%%%%%%%%%%%%%%%%%%%%%%%%%%%%%%%%%%%%%%
%%%%%%%%%%%%%%%%%%%%%%%%%%%%%%%%%%%%%%%%%%%%%%%%%%%%%%%%%%%%%%%%%%%%%%%%%%%%%%

%%%%%%%%%%%%%%%%%%%%%%%%%%%%%%% Figures %%%%%%%%%%%%%%%%%%%%%%%%%%%%%%%%%%%%%%

\vspace{+0.0cm}
%%%%%%%%%%%%%%%%%%%%%%%%%%%%%%%% figure %%%%%%%%%%%%%%%%%%%%%%%%%%%%%%%%%%%%%%
\noindent
\begin{figure}[h]
      \vspace{2.0cm}
   \begin{center}
       \epsfig{file=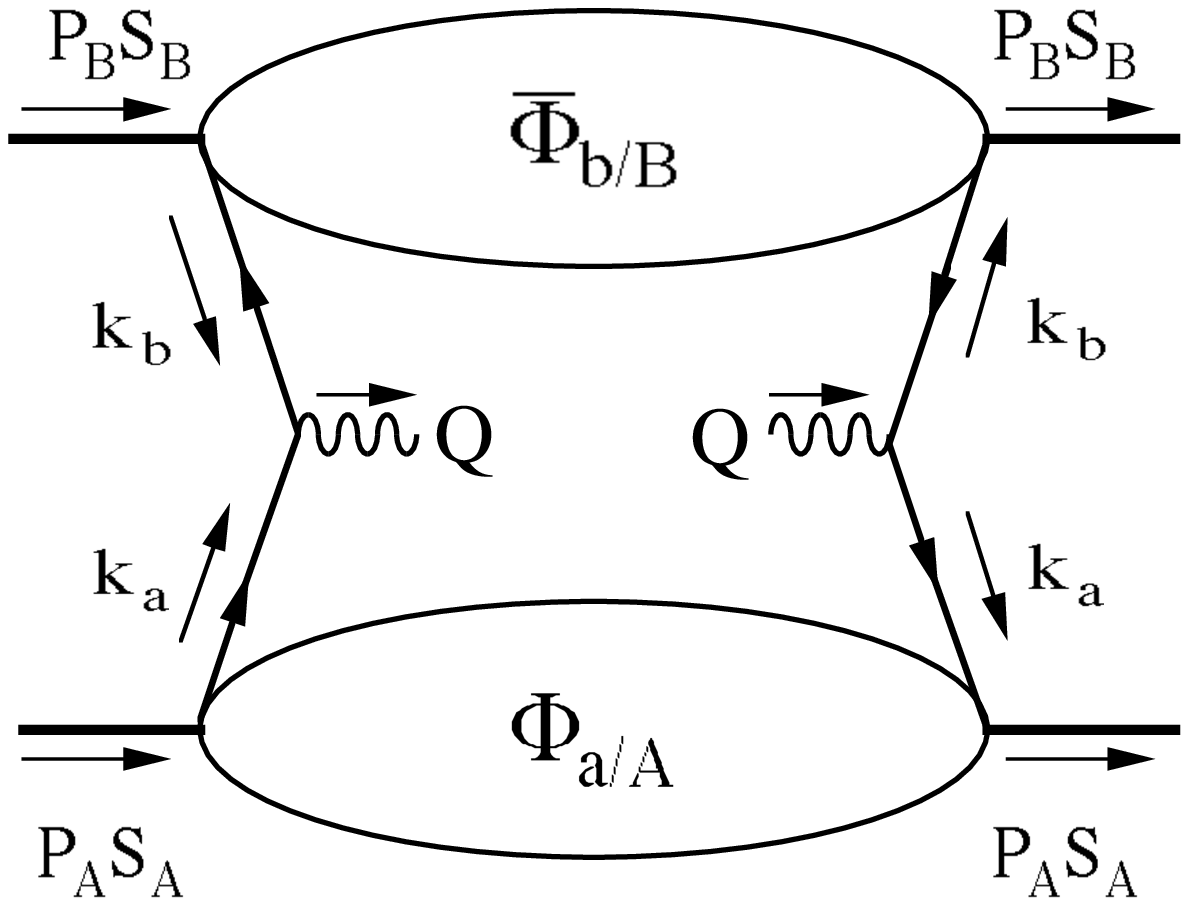,width=8.0cm}
   \end{center}
   \vspace{0.0cm}
   \caption{Drell-Yan process in a parton model.}
   \label{fig:qqbar}
\end{figure}
%%%%%%%%%%%%%%%%%%%%%%%%%%%%%%%% figure %%%%%%%%%%%%%%%%%%%%%%%%%%%%%%%%%%%%%%

\vspace{+1.0cm}
%%%%%%%%%%%%%%%%%%%%%%%%%%%%%%%% figure %%%%%%%%%%%%%%%%%%%%%%%%%%%%%%%%%%%%%%
\noindent
\begin{figure}[h]
   \begin{center}
       \epsfig{file=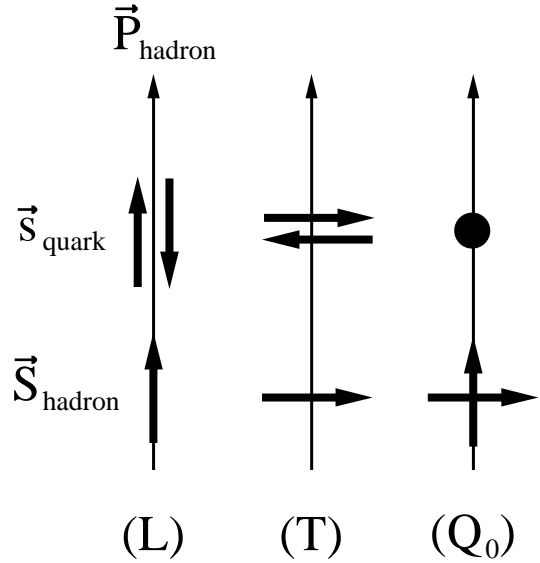,width=8.0cm}
   \end{center}
   \vspace{0.0cm}
   \caption{Polarized parton distributions in a spin-1 hadron
            are illustrated. The figures ($L$), ($T$), and ($Q_0$) indicate
            the longitudinally polarized, transversity, and tensor polarized
            distributions. The mark $\bullet$ indicates the unpolarized
            state.}
   \label{fig:ltq}
\end{figure}
%%%%%%%%%%%%%%%%%%%%%%%%%%%%%%%% figure %%%%%%%%%%%%%%%%%%%%%%%%%%%%%%%%%%%%%%

\end{document}